\documentclass{aa}

\usepackage{txfonts}
\usepackage[dvips]{graphicx}
\usepackage{amsfonts}
\usepackage{amsmath}
\usepackage{amssymb}
\usepackage{natbib}
\usepackage{color,ulem}
\bibpunct{(}{)}{;}{a}{}{,}

\newcommand{\DSct}{$\delta$~Sct}
\newcommand{\GD}{$\gamma$~Dor}
\newcommand{\Teff}{T$_{\rm eff}$}
\newcommand{\lgg}{log\,$g$}
\newcommand{\vsini}{$v$\,sin\,$i$}
\newcommand{\cd}{d$^{-1}$}
\newcommand{\kms}{km\,s$^{-1}$}

\newcommand{\struutup}{\rule{0ex}{3.2ex}}
\newcommand{\struutdown}{\rule[-2ex]{0ex}{2ex}}

\begin{document}
\title{Low-frequency variations of unknown origin in the {\it Kepler} \DSct\ star KIC\,5988140~
= HD\,188774\thanks{Based on data gathered with NASA's Discovery mission {\it Kepler} and with the {\sc Hermes} 
spectrograph, installed at the Mercator Telescope, operated on the island of La Palma by the Flemish Community, 
at the Spanish Observatorio del Roque de los Muchachos of the Instituto de Astrof\'{\i}sica de Canarias, 
and with the 2-m Alfred-Jensch telescope of the Th\"uringer Landessternwarte Tautenburg.},\thanks{Table~A.1 is 
available in electronic form only at the CDS via anonymous ftp to cdsarc.u-strasbg.fr (130.79.128.5)
or via http://cdsweb.u-strasbg.fr/cgi-bin/qcat?J/A+A/.} }
\author{P. Lampens\inst{1} \and A. Tkachenko\inst{2} \and H. Lehmann\inst{3} \and J. Debosscher\inst{2} \and C. Aerts\inst{2,4} \and P. G. Beck\inst{2} \and S. Bloemen\inst{2} \and N. Kochiashvili\inst{5} \and A. Derekas\inst{6} \and J. C. Smith\inst{7} \and P. Tenenbaum\inst{7} \and J. D. Twicken\inst{7}} 
\institute{Koninklijke Sterrenwacht van Belgi\"{e}, Ringlaan 3,
1180, Brussel, Belgium, \email{Patricia.Lampens@oma.be} \and
Instituut voor Sterrenkunde, K.U. Leuven, Celestijnenlaan 200D,
B-3001 Leuven, Belgium \and Th\"{u}ringer Landessternwarte
Tautenburg, 07778 Tautenburg, Germany \and  Department of
Astrophysics, IMAPP, Radboud University Nijmegen, 6500 GL
Nijmegen, The Netherlands \and Abastumani Astrophysical Observatory, 
Ilia State University, 0301 Abastumani, Georgia \and
Konkoly Observatory of the Hungarian Academy of Sciences, PO Box 67, H-1525 Budapest, Hungary \and
SETI Institute/NASA Ames Research Center, Moffett Field, CA 94035}
\date{Received date; accepted date}
\abstract{The NASA exoplanet search mission $Kepler$ is currently providing a wealth of 
light curves of ultra-high quality from space. }{We used high-quality $Kepler$ photometry and 
spectroscopic data to investigate the $Kepler$ target and binary candidate KIC\,5988140. 
We aim to interpret the observed variations of KIC\,5988140 considering three possible 
scenarios: binarity, co-existence of \DSct- and \GD-type oscillations, and rotational 
modulation caused by an asymmetric surface intensity 
distribution. }{We used the spectrum synthesis 
method to derive the fundamental parameters \Teff, \lgg, [M/H], and \vsini\, from the 
newly obtained high-resolution, high S/N spectra. Frequency analyses of both the
photometric and the spectroscopic data were performed. }{The star has a spectral type 
of A7.5~IV-III and a metallicity slightly lower than that of the Sun. Both Fourier 
analyses reveal the same two dominant frequencies F$_1$=2F$_2$=0.688 and F$_2$=0.344~\cd. 
We also detected in the photometry the signal of nine more, significant frequencies 
located in the typical range of \DSct\ pulsation. The light and radial velocity curves 
follow a {similar}, stable double-wave pattern {which are not exactly in anti-phase} 
but show {a relative phase shift of about 0.1 period between the moment of {\it minimum} 
velocity and that of {\it maximum} light.}} 
{Such findings are incompatible with the star being a binary system. {We next show that, 
for all possible (limit) configurations of a spotted surface, the predicted light-to-velocity 
amplitude ratio is almost two orders larger than the observed value, which pleads against 
rotational modulation. The same argument also invalidates the explanation in terms of 
pulsations of type $\gamma$ Dor (i.e. hybrid pulsations). We confirm the occurrence of various 
independent \DSct-type pressure modes in the $Kepler$ light curve. With respect to the low-frequency 
content, however, we argue that the physical cause of the remaining light and radial velocity 
variations of this late A-type star remains unexplained by any of the presently considered scenarios.} }

\titlerunning{Unexplained low-frequency variations in the \DSct~star KIC\,5988140}
\authorrunning{P. Lampens et al.}
\maketitle

\section{Introduction}

The NASA exoplanet search mission {\it Kepler}, launched in 2009,
is currently changing our views about pulsating stars and
asteroseismology thanks to the collection of light curves on all
kinds of variable stars. The nearly continuous time series with
micro-magnitude precision opens up opportunities for high-quality
and in-depth asteroseismic studies with unprecedented detail
\citep{Jenkins2010,Chaplin2011,Bedding2011,Beck2011}. 

A first general characterization of the pulsational behaviour of {750
candidate} A-F type stars with magnitudes between 6 and 15 observed by 
{\it Kepler} has been performed by \citet{Uytterhoeven2011b}. In this
study, 63\% of the sample was assigned to one of three
main classes of pulsating A-F type stars: 27\% are classified as
\DSct\ stars {(206 stars)}, 23\% as hybrid pulsators exhibiting
both \DSct- and \GD-type oscillations (171 stars), and 13\% as \GD\
stars (100 stars). The remaining stars were classified as
rotationally modulated/active stars, binaries, or stars that show
no clear periodic variability. The majority of the hybrid
pulsators shows frequencies with all kinds of periodicities within
the \GD\ and \DSct\ range, which is a challenge for the current
models. \citet{Uytterhoeven2011b} also found indications for the existence 
of \DSct\ and \GD\ stars beyond the edges of the current
{theoretical instability strips, as confirmed in the studies
by \citet{Grigahcene2010} and \citet{Tkachenko2012}}.

Ground-based follow-up observations of the {\it Kepler} targets
have been organized with the goal to obtain precise fundamental 
parameters needed for the seismic modeling of pulsating stars 
\citep[see e.g.][]{Uytterhoeven2011a}. A first sub-sample of A-F
type stars has been recently presented by \citet{Catanzaro2011}.

We focus here on the {\it Kepler} target KIC\,5988140 (HD\,188774, 
{\it Kepler} magnitude of 8.852). In the study by \citet{Uytterhoeven2011b}, this 
late A/early F-type star is classified as a binary with \DSct\
pulsations. At least 12\% of the entire sample was identified as
belonging to a binary or a multiple system. We further remark that
its light curve resembles that of KIC\,9664869 \citep[see
panel c of Fig.~5 in][]{Uytterhoeven2011b} which is in turn classified as a
hybrid pulsator of the middle type, i.e. neither of dominating
\DSct\ type nor of dominating \GD\ type. \citet{Catanzaro2011}
investigated KIC\,5988140 based on low and medium resolution
spectroscopic data and reported the fundamental {atmospheric 
parameters shown in Table~\ref{Table:AtmosphericParameters}}. 
{Unlike \citet{Uytterhoeven2011b}, these authors concluded 
that KIC\,5988140 belongs to the hybrid \DSct\ - \GD\ class.}{ In this 
study, we use new high-resolution spectra of KIC\,5988140 to investigate 
its radial velocity (RV) curve and to re-determine its fundamental 
parameters, with the goal to identify the possible cause(s) of the 
observed photometric and spectroscopic variations.}

\section{Observations}\label{Section:Obs}

Our analysis is based on high-quality photometric and
spectroscopic data. The light curves have been gathered by the
{\it Kepler} satellite in the so-called {\it Long Cadence} mode
(with a time resolution of 29.4 min). The {\it Kepler} data are
made available in quarters, i.e. periods between two spacecraft
rolls. The {(uncorrected)} data of different quarters have different zero-points
and suffer from instrumental drift which may vary from quarter to
quarter. To correct for these, we used a package of Fortran
routines (kindly made available by L.~Balona), which removes the
jumps between the quarters and the drifts in an automatic way. The
processed data represent normalized relative intensities expressed
in parts per thousand (ppt). Using this procedure may affect some of the
lowest frequencies found in a period-search analysis.
\citet{Balona2011} tested this assumption and concluded that the
frequencies above 0.1~\cd~are not affected, {while the amplitudes 
of those below 0.1~\cd~may change slightly}. For this case study, we 
analysed the {\it Kepler} data from seven different quarters, i.e. 
{Q0-Q2,Q4-Q6,Q8}, which resulted in 22\,234 measurements in total and 
from which we extracted 20\,668 useful measurements. {The Nyquist 
frequency is 24.47~\cd~and the frequency resolution is 0.0022~\cd~(for 
the time base of $682$ days)~\citep{Loumos1978}. Two small aliasing peaks 
located at respectively 0.0033 and 0.0052~\cd, and caused by the missing 
quarters Q3 and Q7, can be identified in the corresponding spectral window.} 

We also acquired a series of high-resolution spectra in two consecutive
years. In 2010, the spectra were taken with the {\sc Hermes}
spectrograph attached to the 1.2-m Mercator telescope
(Observatorio del Roque de los Muchachos, La Palma, Canary
Island). {\sc Hermes} (High Efficiency and Resolution Mercator Echelle
Spectrograph) is a fibre-fed spectrograph which samples the entire
optical range from 380 to 900~nm with a resolution of 85\,000
\citep{Raskin2011}. The spectra acquired in 2011 have been taken
with the Coud\'e-Echelle spectrograph attached to the 2-m
telescope at the Th\"{u}ringer Landessternwarte Tautenburg. The
Tautenburg spectra have a spectral resolution of 32\,000 and cover
the wavelength range from 472 to 740~nm. {Exposure times of the same order 
as the temporal resolution of the $Kepler$ data were used in both cases (i.e. 
between 6 and 30 min depending on the conditions).}

The data have been reduced {using the dedicated pipeline for {\sc Hermes} 
spectra and standard ESO-MIDAS packages for the extracted, merged {\sc Hermes} 
spectra and the Tautenburg spectra, respectively}. The data reduction included 
bias and stray-light subtraction, cosmic rays filtering, flat fielding, wavelength
calibration by {ThArNe} lamps, order merging, and normalization to
the local continuum. All the spectra were additionally corrected in
wavelength for individual instrumental shifts by using a large
number of telluric O$_2$ lines. The cross-correlation technique
was used to estimate radial velocities (RVs) of the individual
spectra. All the spectra have been 
shifted in wavelength according to their derived RVs and {co-}added 
to build {two} high(er) signal-to-noise ratio (S/N) 
{mean spectra}. {The {\sc Hermes} (respectively Tautenburg) mean spectrum 
was next analysed individually}. {Since both analyses gave consistent results, 
and because the {\sc Hermes} mean spectrum covers a wider spectral range and is 
of excellent quality (S/N = 180-200 at $\lambda$ $\sim$ 500~nm), we will refer 
to the latter (only)} as 'the mean spectrum' in the following sections.

The journal of observations is given in Table~\ref{Table:observations}. 
The associated errors quoted in Table~\ref{Table:observations} are the 1-$\sigma$ 
formal errors obtained during the fitting process. 

\section{Spectroscopy}

\subsection{Spectrum analysis}\label{Section:SpectrumAnalysis}

\begin{table}
\tabcolsep 0.8mm\caption{\small Fundamental parameters {and the 1-$\sigma$ formal errors} as derived
from the mean spectrum. The KIC parameters {and their typical errors (in parentheses)} are listed 
for comparison.}
\begin{tabular}{llcccc}
\hline\hline \multicolumn{1}{c}{[M/H]\rule{0pt}{9pt}} &
  \multicolumn{1}{c}{\Teff ($K$)} & \multicolumn{1}{c}{\lgg} &
  \multicolumn{1}{c}{\vsini\, } & \multicolumn{1}{c}{$\xi$\, } & 
  \multicolumn{1}{c}{SpT}\\
\hline \multicolumn{3}{c}{} &
  \multicolumn{1}{c}{(\kms)} & \multicolumn{1}{c}{(\kms)} & 
  \multicolumn{1}{c}{}\\
\hline \multicolumn{6}{c}{{this work}\rule{0pt}{15pt}}\\
--0.30 {\tiny (0.05)}\rule{0pt}{9pt} & 7600 {\tiny (30)} &
3.39 {\tiny (0.12)} & 52.0 {\tiny (1.5)} & 
3.16 {\tiny (0.20)} & A7.5\,IV-III\\
\multicolumn{6}{c}{{KIC}\rule{0pt}{15pt}}\\
--0.54 {\tiny (0.50)}\rule{0pt}{9pt} & 7451 {\tiny (200)} &
3.54 {\tiny (0.50)} & --- & --- & A8\,IV-III\vspace{1.5mm}\\
\multicolumn{6}{c}{{Catanzaro et al.\, (2011)}\rule{0pt}{15pt}}\\
--- & 7400 {\tiny (150)} & 3.7 {\tiny (0.3)} & --- & --- 
& --- \vspace{1.5mm}\\
\hline
\end{tabular}
\label{Table:AtmosphericParameters}
\end{table}

\begin{table}
\tabcolsep 1.5mm\caption{\small Abundances for individual elements sorted according to 
their {uncertainties}. N is the number of lines with a line depth larger than rms.
The {uncertainties} quoted between parentheses are given in terms of last digits.}
\begin{tabular}{lrrrrrr}
\hline\hline \multicolumn{1}{c}{Element\rule{0pt}{9pt}} &
\multicolumn{1}{c}{N} & \multicolumn{1}{c}{N~{\small I}} & \multicolumn{1}{c}{N~{\small II}} & \multicolumn{1}{c}{solar} & \multicolumn{1}{c}{[A/H]} & \multicolumn{1}{c}{dA}\\
\hline Fe\rule{0pt}{9pt} & 861 & 753 & 108 & --4.59 & --4.99(05) & --0.40(05)\\
Ca & 55 & 42 & 13 & --5.73 & --5.88(05) & --0.15(05)\\
Ti & 162 & 23 & 139 & --7.14 & --7.42(09) & --0.28(09)\\
Cr & 149 & 71 & 78 & --6.40 & --6.60(11) & --0.20(11)\\
Mg & 32 & 25 & 7 & --4.51 & --4.89(15) & --0.38(15)\\
Ni & 80 & 74 & 6 & --5.81 & --6.05(16) & --0.24(16)\\
Sc & 24 & 0 & 24 & --8.99 & --9.17(19) & --0.18(19)\\
Mn & 42 & 35 & 7 & --6.65 & --7.07(21) & --0.42(21)\\
V & 44 & 2 & 42 & --8.04 & --8.27(23) & --0.23(23)\\
C & 14 & 14 & 0 & --3.65 & --3.83(24) & --0.18(24)\\
Y & 28 & 0 & 28 & --9.83 & --10.01(25) & --0.18(25)\\
Si & 13 & 4 & 9 & --4.53 & --4.77(31) & --0.24(31)\\
Co & 14 & 13 & 1 & --7.12 & --7.43(36) & --0.31(36)\\
Sr & 4 & 0 & 4 & --9.12 & --9.02(42) & +0.10(42)\\
Ba & 7 & 0 & 7 & --9.87 & --9.57(46) & +0.30(46)\\
Eu & 6 & 0 & 6 & --11.52 & --11.47(49) & +0.05(49)\\
\hline
\end{tabular}
\\\vspace{0.3mm}\\
{\footnotesize {\it Note}: Columns 1 to 4 represent the element designation, the total 
number of spectral lines in the considered wavelength region, and in the first and the 
second ionization stage, respectively, while columns 5 to 7 list the solar composition, 
the individual abundances, and the deviation of those with respect to the solar 
values \citep{Asplund2009}.} 
\label{Table:Abundances}
\end{table}

The mean spectrum of KIC\,5988140 was analysed by means of 
the GSSP programme package \citep{Tkachenko2012} which is based 
on the method of synthetic spectra. The idea is to use a large, dense 
enough grid of pre-computed theoretical spectra and perform a
spectrum-by-spectrum comparison with the observations until a
global minimum {of $\chi^2$} in parameter space is found. In the first step,
synthetic spectra are computed on a grid of effective
temperature \Teff, surface gravity \lgg, projected
rotational velocity \vsini, microturbulent velocity $\xi$, and
metallicity [M/H]. The derived metallicity is used as an
initial guess for the chemical composition of the star and
individual elemental abundances are iterated together with the
other four fundamental parameters in the second step. The measurement 
{uncertainties} of \Teff, \lgg, \vsini, $\xi$, [M/H], and the
individual abundances are determined as 1-$\sigma$ confidence levels
obtained from $\chi^2$-statistics. For the calculation of
synthetic spectra, we used the code SynthV \citep{Tsymbal1996} based
on the local thermal equilibrium (LTE) theory whereas atmosphere models 
have been computed with the most recent, parallelized version of the 
LLmodels \citep{Shulyak2004}. All synthetic spectra were 
computed in the 3800--5700~\AA\, wavelength region which is almost 
free of telluric contributions but includes several hydrogen lines of 
the Balmer series and a large number of metal lines. A more detailed
description of the method and its application to the spectra of
{\it Kepler} $\beta$\,Cep and SPB candidate stars as well as \DSct~
and $\gamma$\,Dor candidate stars are given by \citet{Lehmann2011}
and by \citet{Tkachenko2012}, respectively.

Table~\ref{Table:AtmosphericParameters} gives the results of the
spectrum analysis in comparison with the {\it Kepler Input
Catalog} (KIC) values. The {errors} of the KIC parameters
are typical values, i.e. $\pm$200~K for the effective temperature and 
$\pm$0.5~dex for both the surface gravity and the metallicity. The 
derived fundamental parameters agree within the (KIC) errors, 
though there is a rather large difference in the value of [M/H]
between ours and the KIC one. {The new fundamental parameters
also agree} with respect to the values reported by \citet{Catanzaro2011}.
Spectral types and the luminosity classes listed in Table~\ref{Table:AtmosphericParameters} 
were derived using interpolation in the tables published by
\citet{Schmidt-Kaler1982}.

Table~\ref{Table:Abundances} lists the abundances of individual elements as
determined from the mean spectrum. 
Apart from Sr, Ba, and Eu, the abundances of all elements are compatible with 
the derived metallicity of the star. The three elements mentioned above show
considerable overabundances {but their uncertainties} are also large.

\begin{figure}
\centering
\includegraphics[scale=0.90]{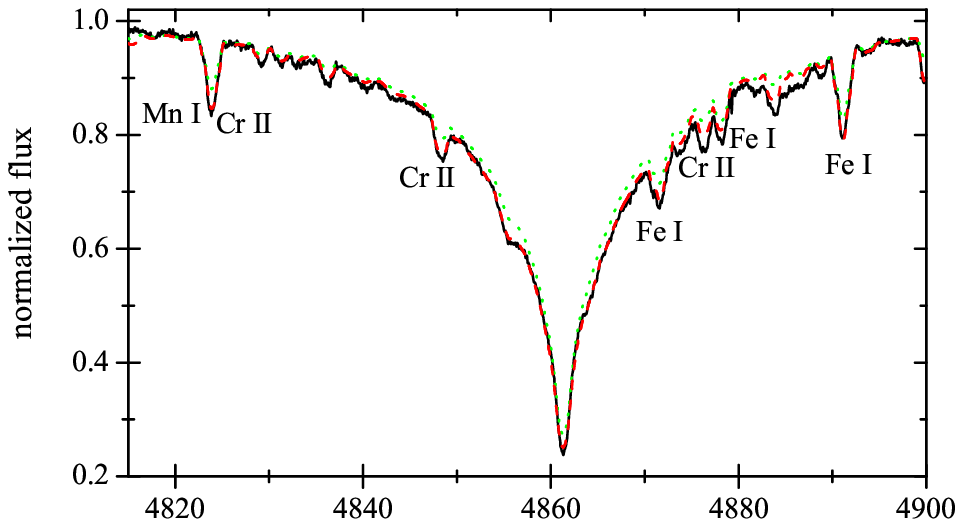}\vspace{2mm}
\includegraphics[scale=0.90]{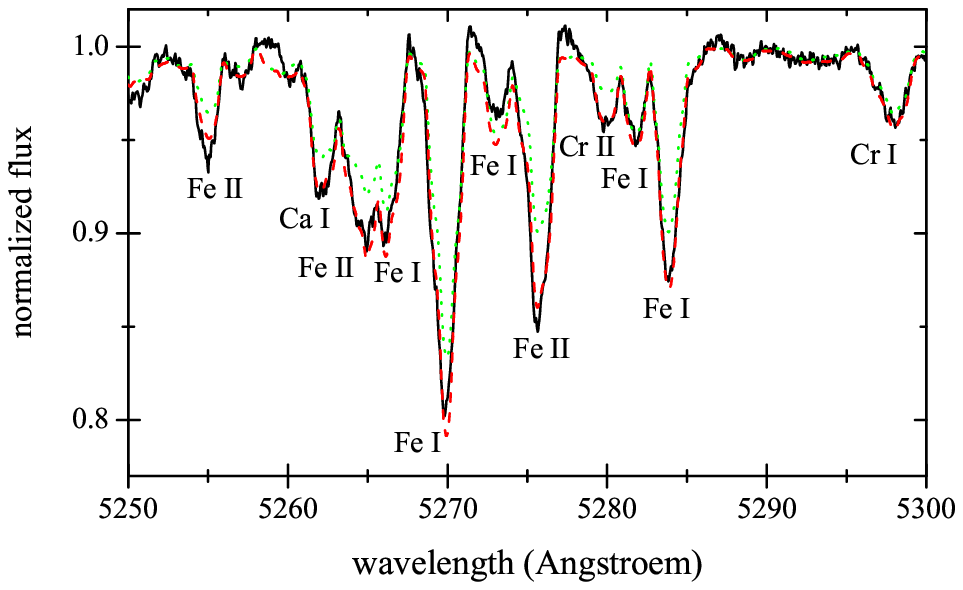}
\caption{{\small Fit of the observed spectrum of KIC\,5988140
(black) in two different wavelength regions assuming
T$_{\rm{eff}}$ = 7600/7400~K, log\,$g$ = 3.4/3.5 and [M/H] =
--0.3/--0.5~dex (resp. dashed red and dotted green lines). A colour 
plot is provided in the online version.} }
\label{KIC05988140_comp}
\end{figure}

\begin{figure}
\centering
\includegraphics[angle=0,scale=0.37,viewport= 04 08 635 525]{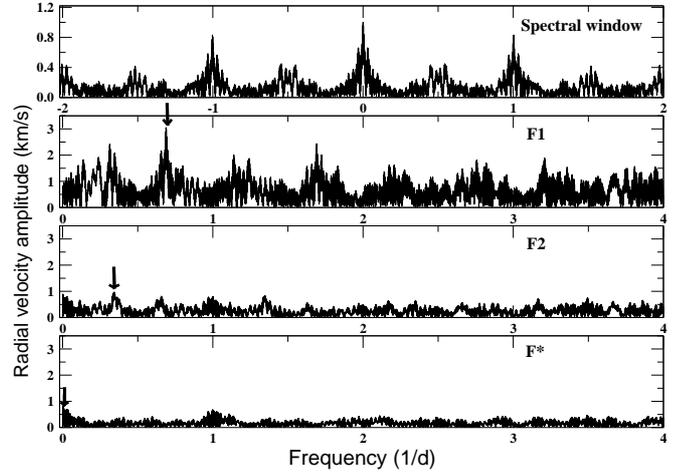}
\caption{{\small Successive periodograms of the ground-based RV
data of KIC\,5988140 (top: spectral window, bottom: after prewhitening for 
F$_1$ and F$_2$).  } }
\label{KIC05988140_FOU}
\end{figure}

Figure~\ref{KIC05988140_comp} shows {parts of} the observed spectrum (black line)
in comparison with the two synthetic spectra computed respectively
based on the parameters derived by us (dashed line) and on those listed in
the KIC (dotted line). Obviously, the newly derived fundamental parameters provide a 
better fit of the observed spectrum in both illustrated wavelength 
regions. The same is true for the majority of the other metal lines found 
in the considered wavelength range.

\subsection{Radial velocity variations}\label{Section:RVs}

The RVs measured (by cross-correlation, see Sect.~\ref{Section:Obs}) from all 
recently observed spectra (i.e. 40 measurements in total) were searched for 
possible periodicities by means of the {\sc Period04} programme \citep{Lenz2005}.
Figure~\ref{KIC05988140_FOU} illustrates the Fourier spectrum of
the RV data set. {The corresponding spectral window 
reveals the presence of {the 1 year$^{-1}$ as well as} 
80\% high 1 day$^{-1}$ alias peaks.}
The {\it dominant} period P$_1$ is detected at 1.4534$\pm$0.0085~d. 
After prewhitening with the corresponding frequency $F_1$=0.6880$\pm$0.0038~\cd, 
a Fourier analysis of the residuals reveals a {\it second} significant
frequency $F_2$=0.3440$\pm$0.0002~\cd~corresponding to the
period P$_2$=2.9070$\pm$0.0002~d (see Table~\ref{Tab:freq_RVs}).
{We computed 100 Monte Carlo simulations to estimate the errors 
on the frequencies and amplitudes (cf. {\sc Period04}).}
{This second frequency appears quite significantly after prewhitening, 
and is not affected by the aliasing peaks located at $\sim$0.5~\cd~in the 
spectral window (as will be confirmed in Sect.~\ref{Section:Kplr}). We 
also remark that the ratio between both frequencies is exactly 2.}
{The third frequency, $F_*$ = 0.0013~\cd, is apparently caused 
by a small shift in zero-point between the RVs originating 
from two different instruments.} {We kept its value fixed during the
Monte Carlo error computations.}{ The tabulated amplitude-to-noise 
ratios much larger than 4.0 \citep{Breger1993} show that all three 
frequencies are significant.}{ The fraction of the variance removed 
after prewhitening with three frequencies, ($1 - R$), equals 87\%.}

\begin{figure}
\centering
\resizebox{8cm}{!}{\includegraphics[angle=-90,scale=0.30]{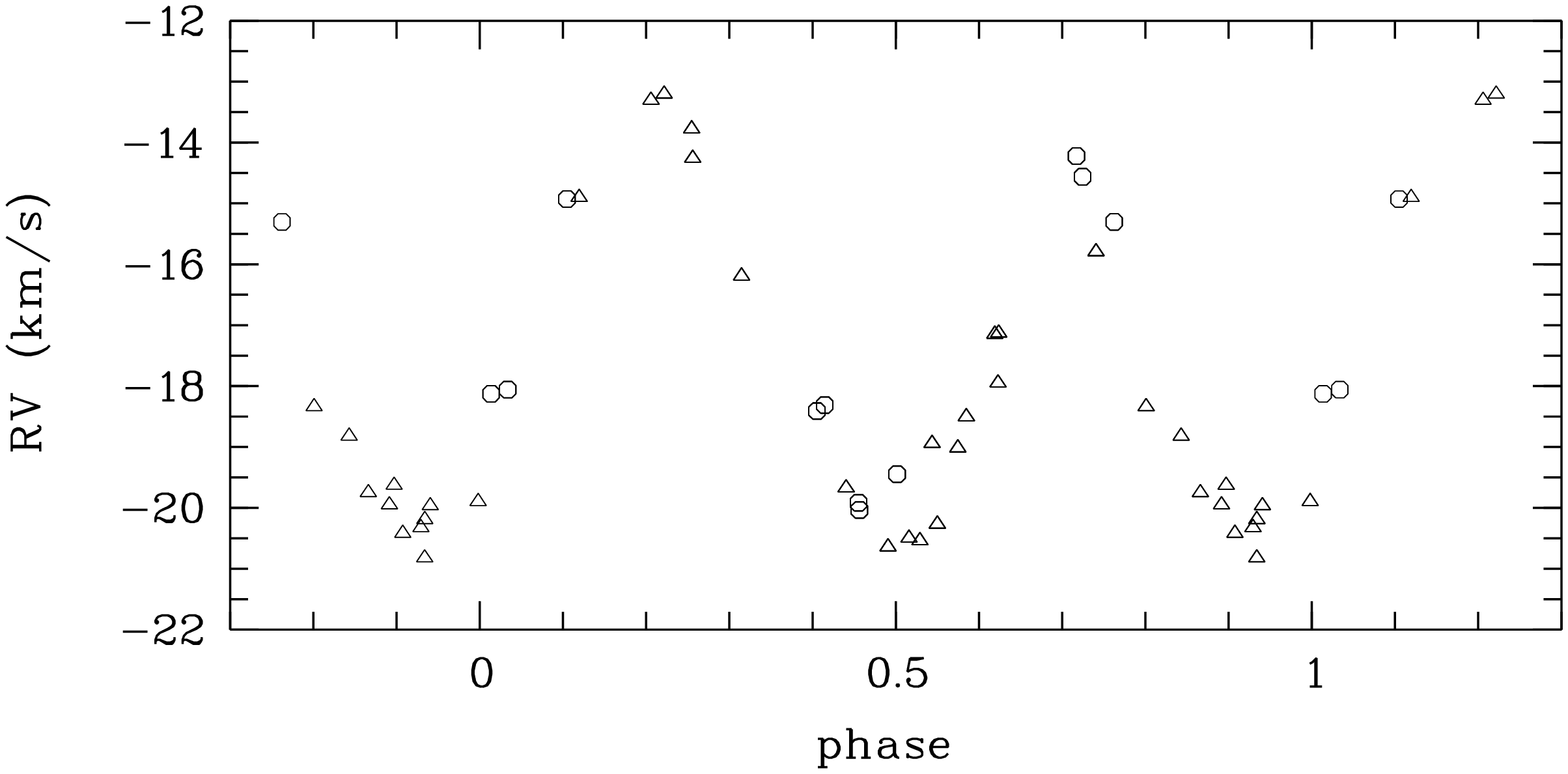}}
\resizebox{7.5cm}{!}{\includegraphics[angle=0,scale=0.30]{Lampensp_fig5.eps}}
\caption{{\small {\it Top:} RVs of KIC\,5988140 folded on the
period P$_2\sim$2.91~d. {\sc Hermes} RVs are indicated by
triangles and TLS ones by open circles. {\it Bottom:} Portion of 
the {\it Kepler} light curve illustrating the near 2-hr period oscillations.
The X-axis shows both BJD and phase. {In both panels, phase zero 
refers to (B)JD~2455087.62.}}}
\label{KIC05988140_RV_LC}
\end{figure}

Figure~\ref{KIC05988140_RV_LC} illustrates the RV curve based on
the Tautenburg and {\sc Hermes} spectra, phased on the period
P$_2$ (top), as well as a small portion of the {\it Kepler} light curve
(bottom). Phase zero in the RV curve corresponds to the time of the 
highest maximum observed in this part of the light curve which occurs
at (B)JD~2455087.62. {The following ephemeris has been adopted:
\[ (B)JD_{max\_light/RV} =  (B)JD~2455087.62 + E \times 2.9071~d. \]  }
The RVs thus folded follow a clear double-wave pattern (two maxima, 
two minima). Apart from its obvious double-wave shape, the mean radial 
velocity curve appears slightly {anharmonic} (the phase difference between 
the {first two periodic signals equals 0.34, see Table~\ref{Tab:freq_RVs})}. 

\begin{table}[t]
\caption[]{Frequencies detected in the RV data of KIC\,5988140}
\begin{tabular}{lccccc}
\hline
ID  & Frequency    & Amplit. & Phase  & S/N & Note \struutup\\
    & ($\pm$ error) &  ($\pm$ 0.05)   &       &     &  \\
    & \cd        &   km/s   & 2$\pi$rad  &     &   \struutdown\\
\hline
$F_1$    &  0.6880 (0.0038) & 2.93 &  0.57 & 33 & = 2$F_2$\struutup\\
$F_2$    &  0.3440 (0.0002) & 1.34 &  0.91 & 15 & \\
$F_*$    &  0.0013 (fixed)  & 1.41 &  0.04 & 15 & \struutdown\\
\hline
\vspace{1mm}
\end{tabular}
\label{Tab:freq_RVs}
\end{table}

\subsection{Least-squares deconvolution}

\begin{figure}
\centering
\includegraphics[scale=0.90]{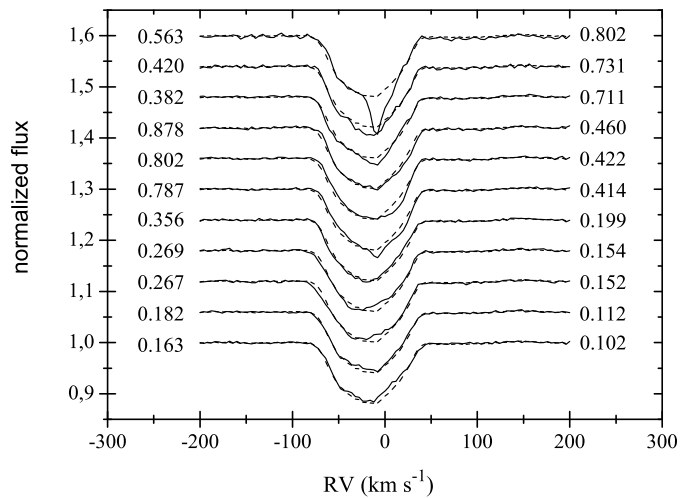}
\caption{{\small LSD-profiles computed based on the individual
(solid) and mean (dashed) spectra.}} \label{KIC05988140_LSD}
\end{figure}

A fit of the mean, observed spectrum of KIC\,5988140 by the
synthetic one provides quite smooth residuals indicating that this
is 
not a double-lined spectroscopic binary. High S/N averaged profiles 
computed by means of the least-squares deconvolution (LSD) technique 
\citep{Donati1997} confirm this finding.
Figure~\ref{KIC05988140_LSD} illustrates the profiles computed
from 11 Tautenburg spectra (solid) which are shifted in Y-axis for
clarity. The phases indicated on the plot to the left and to the
right of the profiles were computed assuming two different
periods, P$_1$ and P$_2$, respectively, and the time of minimum
corresponding to the deepest minimum in the {\it Kepler} light curve
(at BJD$\sim$2455085.50, cf. Figure~\ref{KIC05988140_RV_LC}, bottom 
part). The LSD-profile computed from the mean spectrum which has been 
used for the detailed analysis in Section~\ref{Section:SpectrumAnalysis} 
is shown by a dashed line for comparison. Obviously, there is no {indication} 
of a secondary component in the spectra. The first profile from the top 
appears to be highly asymmetric and has a shape typical of that caused by
the Rossiter--McLaughlin effect \citep{McLau1924,Ross1924}. However, the 
computed phases indicate that the corresponding spectrum has been taken at 
the phase {near} a time of {secondary} maximum light in the 
{\it Kepler} light curve and not during the eclipse phase. This fact allows 
to exclude eclipses as a possible explanation for the observed anomalous
shape of spectral lines at this phase, thus the nature of the
anomaly remains unclear. On top of that, there are also obvious
line profile variations occurring on a short time scale which might
be caused by stellar oscillations (cf. Sect.~\ref{Section:Kplr}).

\section{{\it Kepler} photometry}\label{Section:Kplr}

\begin{figure}
\includegraphics[angle=-90,scale=0.33]{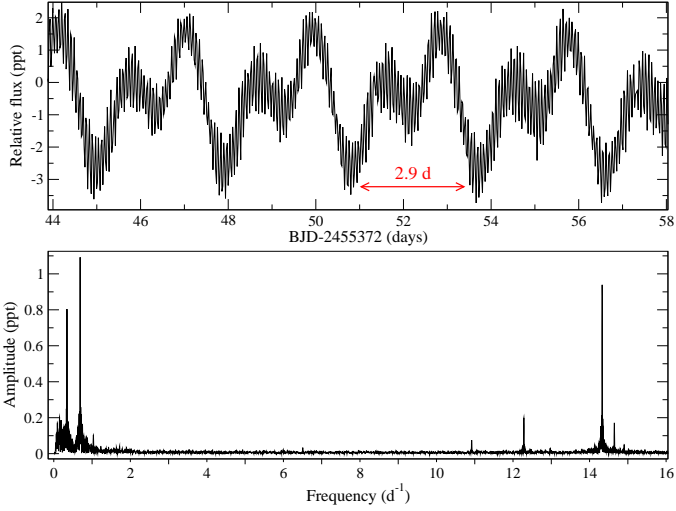}
\caption{{\small Part of the Q6-quarter of the {\it Kepler} light curve of KIC\,5988140~
(top) and amplitude spectrum of the entire Q6-quarter data string (bottom). }} \label{KIC05988140_LC}
\end{figure}

Similar to the analysis of the spectroscopic data, a search for 
periodicities has been performed using the {\sc Period04} programme. 
Figure~\ref{KIC05988140_LC} represents a 2-weeks long string of the 
Q6 {\it Kepler} light curve (top) together with the amplitude spectrum 
based on {the entire Q6-quarter data string} (bottom). The 
highest peak appears at the frequency of 0.687993~\cd~(P$_1$ = 1.45350~d),
followed by the frequency of 0.343984~\cd~(P$_2$ = 2.90711~d). This
result is entirely consistent with the results from the Fourier
analysis of the RVs (cf. Table~\ref{Tab:freq_RVs}).
{Note that the phase difference between the first two periodic 
signals is 0.52, therefore not exactly matching the one from the radial 
velocity curve. The ratio $F_1$/$F_2$ equals the factor 2, which 
could in principle be explained as tidal modulation in a close binary system 
(e.g. in ellipsoidal systems), but this explanation is not corroborated 
by the facts that the same factor appears in the frequency-analysis of the RVs 
nor that the total RV amplitude is very low.} Besides the peaks corresponding 
to the periods P$_1$ and P$_2$, the frequency analysis {of the complete data 
set} revealed the existence of at least twenty-five more frequencies, with a 
significance above ten times the noise level in Fourier space. While most of 
these are located $<$ 0.1~\cd~indicating either long-term changes or remaining 
instrumental effects (see cautionary note from Sect.~\ref{Section:Obs}), 
some are obviously harmonics of P$_2$: apart from P$_1$=P$_2$/2, we also 
{find} P$_2$/3, P$_2$/4, and P$_2$/5. However, more significant is the 
detection of nine frequencies located beyond 10~\cd: they all appear to lie 
in the typical $\delta$\,Sct range (see Table~\ref{Tab:Freq_Phot}). 
{As before, the errors on the frequencies and amplitudes were computed
based on Monte Carlo simulations, except for $F_1$ which was kept fixed 
with respect to $F_2$ (cf. {\sc Period04}).}
{The fraction of the variance removed after prewhitening with 27 frequencies 
(i.e. after removal of the 27 first frequencies), ($1 - R$), equals 99\%.}
Figure~\ref{KIC05988140_RV_LC} (bottom) shows a randomly selected
light curve portion illustrating the short-period variations over
one and a half full cycle with respect to P$_2$. Besides the fact 
that the light curve appears to be {lagging in phase behind} the 
RV-curve folded {on} P$_2$ {by at least 0.2 period (i.e. maximum 
to maximum)}, a very clear beating of the most significant frequency located 
at 14.32981~\cd~(corresponding to a pulsation period of 1h40m) can be seen, 
caused by the presence of six more frequencies located between 14 and 15~\cd. 
Among these, {a few frequency spacings} are repeatedly occurring. For example, 
the frequencies $F_3$, $F_7$ and $F_{23}$ are separated by almost exactly 
0.040~\cd. {The couples ($F_5$,$F_9$) and ($F_{19}$,$F_5$) show a difference 
of 0.31 and 0.25 \cd, respectively.} Two other \DSct-type frequencies are 
$F_4$ = 12.28328~\cd~and $F_{14}$ = 10.91996~\cd~(with respective period ratios 
of 0.86 and 0.76). The latter is an interesting period ratio, close to the ratio 
1H/F expected for radial modes in such a star. The occurrence of $F_{17}$ = 23.89006~\cd, 
on the other hand, is questionable, since it lies very close to the Nyquist
frequency of the data set. We further notice the presence of
various long-term periodicities (frequencies $<$ 0.01~\cd) which
appear with a low significance, though none as small as the
frequency of 0.001~\cd~found below the detection limit in the
RV data (cf. Sect.~\ref{Section:RVs}). We also 
recall that such {very low frequencies} might be affected by the data 
(pre)processing algorithms. 

Next, we tested the stability of the properties of the \DSct-type
pulsations by searching for a best-fit model allowing for a
periodic variation of the phase shifts of these frequencies in the
multi-parameter solution. For this, we used the model called 'Periodic 
Time Shift' in {\sc Period04} (cf. PTS model under expert mode). 
The computations show that no better fit can be found by including a 
possible (long-term) variation of the phase shifts which might possibly 
be due to a modulation by either one of the low frequencies P$_1$ or P$_2$.
Our conclusion is that the properties of the \DSct-type frequencies are 
stable.

\begin{table}[t]
\caption[]{Over 25 frequencies detected in the photometry of
KIC\,5988140. }
\begin{tabular}{lcccc}
\hline
ID  & Frequency    & Amplit. & S/N & Note \struutup\\
    & ($\pm$ error) &  ($\pm$ 0.001)   &     &  \\
    & \cd        &   ppt    & ppt       &   \struutdown\\
\hline
$F_1$    &  0.687993 (fixed)  & 1.261 &  411 &  = 2$F_2$\struutup\\
$F_2$    &  0.3439842 (3E-07)  & 1.123 &  335 & \\
$F_3$    &  14.32981 (2E-06)  & 0.910 & 1207 &  \\
$F_4$    &  12.28328 (4E-06)  & 0.192 &  256 &  \\
$F_5$    &  14.64803 (5E-06)  & 0.146 &  230 &  \\
$F_7$    &  14.28878 (9E-06)  & 0.094 &  123 &  \\
$F_9$    &  14.33072 (2E-05)  & 0.084 &  112 &  \\
$F_{10}$   &  1.031966 (1E-05)  & 0.065 &   33 &  = 3$F_2$\\
$F_{13}$   &  14.12919 (1E-05)  & 0.055 &   70 &  \\
$F_{14}$   &  10.91996 (2E-05)  & 0.051 &  129 &  \\
$F_{17}$   &  23.89006 (2E-05)  & 0.045 &   49 &  \\
$F_{18}$   &  1.375905 (2E-05)  & 0.044 &   38 &  = 4$F_2$\\
$F_{19}$   &  14.90637 (2E-05)  & 0.043 &   77 &  \\
$F_{23}$   &  14.24855 (2E-05)  & 0.033 &   43 &  \\
$F_{28}$   &  1.719992 (3E-05)  & 0.028 &   33 &  = 5$F_2$\struutdown\\
\cline{1-5}\\
$F_6^{*}$    &  0.057054 (1E-05)  & 0.156 &   45 &  \\
$F_8^{*}$    &  0.053534 (1E-05)  & 0.092 &   26 &  \\
$F_{11}^{*}$   &  0.060134 (2E-05)  & 0.094 &   27 &  \\
$F_{12}^{*}$   &  0.090642 (1E-05)  & 0.066 &   19 &  \\
$F_{15}^{*}$   &  0.075828 (1E-05)  & 0.065 &   19 &  \\
$F_{16}^{*}$   &  0.083162 (2E-05)  & 0.049 &   14 &  \\
$F_{20}^{*}$   &  0.004253 (2E-05)  & 0.044 &   13 &  \\
$F_{21}^{*}$   &  0.033734 (2E-05)  & 0.036 &   20 &  \\
$F_{22}^{*}$   &  0.062921 (2E-05)  & 0.048 &   14 &  \\
$F_{24}^{*}$   &  0.050308 (2E-05)  & 0.042 &   12 &  \\
$F_{25}^{*}$   &  0.092842 (2E-05)  & 0.036 &   10 &  \\
$F_{26}^{*}$   &  0.070695 (3E-05)  & 0.035 &   10 &  \\
$F_{27}^{*}$   &  0.110442 (3E-05)  & 0.029 &    8 &\struutdown\\
\hline
\end{tabular}
\\\vspace{0.2mm}\\
{\footnotesize$ ^{*}$: Frequencies not considered as meaningful, see Sect.~\ref{Section:Obs}.} 
\label{Tab:Freq_Phot}
\end{table}

\begin{figure}[t]
\begin{center}
\includegraphics[angle=0,scale=0.32]{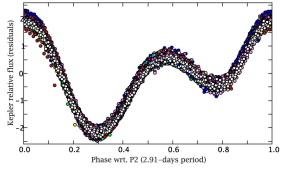}
\caption{{\small ``Basic'' light curve of KIC\,5988140 based on all the data and 
folded on the period P$_2$ {(after prewhitening for 23 frequencies)}. }}
\label{K005988140_Onlyf0f1}
\end{center}
\end{figure}

\begin{figure}
\includegraphics[angle=0,scale=0.30]{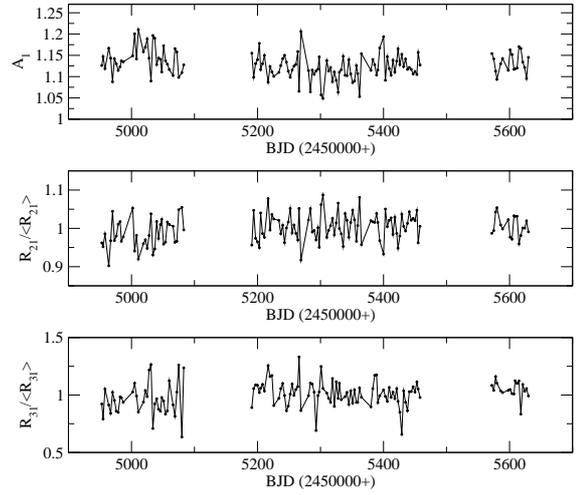}
\caption{{\small Temporal evolution of the (relative) amplitudes $A_{1}$ and  
$R_{21},\ R_{31}$ of the ``basic'' light curve of KIC\,5988140. }}
\label{fourier_parameters}
\end{figure}

We also verified the stability of the shape of the ``basic'' light
curve (i.e. cleaned for the near 2hr-period variations and the
long-term trends) based on the main frequency {and the harmonics} by 
prewhitening with all other frequencies detected during the
Fourier analysis, and by plotting the folded residual light curve
using colours for different subsets.
Figure~\ref{K005988140_Onlyf0f1} shows the resulting phase diagram
when subdividing the data into bins of tens of days. {We see
no obvious temporal change in the shape of this curve. As an independent test, 
we divided the ``basic''  {\it Kepler} light curve into different 
cycles and fitted each with a 4th-order trigonometric polynomial according to 
\begin{equation} m=A_{0}+\sum_{j=1}^{4} A_{j} \cdot \sin\left(2\pi \cdot jF_{2} \cdot t + \phi_{j}\right), \end{equation} 
where {\it m} is the magnitude, {\it A$_{j}$} is the amplitude, {\it jF$_{2}$} stands for 
the frequency $F_{2}$ and its harmonics, {\it t} is the time of the observation, 
$\phi_{j}$ is the phase and index {\it j} runs from 1 to 4. Then, we characterised 
each cycle by its Fourier coefficients \citep{sim82}. In Fig.~\ref{fourier_parameters}, 
we represent the amplitude, $A_{1}$, and the relative variations of the parameters 
$R_{21}=A_{2}/A_{1},\ R_{31}=A_{3}/A_{1}$ as a function of time. The amplitude $A_{1}$ 
shows a scatter of about 0.08~ppt which is ten times larger than the average fit error 
($\sim$0.008~ppt). While the relative scatter of $R_{21}$ is small, the relative scatter 
of $R_{31}$ is more important, but considering that the amplitude $A_{3}$ is much smaller, 
than $A_{1}$, the corresponding variation remains insignificant and we find no 
clear trend. This demonstrates that the ``basic'' light curve 
is stable over a period of almost two years ($T = 682$ days). 
 }

{Considering that the time resolution in {\it Long Cadence} mode
is of the order of 30 min, we must caution the determination of the
amplitudes of the \DSct-type frequencies in Table~\ref{Tab:Freq_Phot}. 
Whereas the amplitudes in the low-frequency regime may be correctly 
evaluated, the ones in the high-frequency regime might actually be 
slightly underestimated.}

{Making use of Figures~\ref{KIC05988140_RV_LC} and ~\ref{K005988140_Onlyf0f1}, we can 
also verify the phase relationship} between the ``basic'' light and RV 
curves: both curves are {not exactly in anti-phase with respect to each 
other: the moments of {\it minimum} RV precede the moments of 
{\it maximum} light by about 0.1 period.}  

\begin{figure}
\begin{center}
\includegraphics[angle=0,scale=0.30]{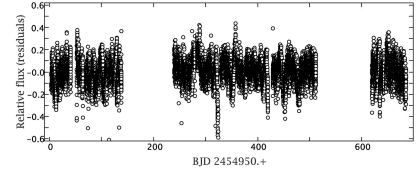}
\caption{{\small Residual light curve of KIC\,5988140 based on
all the data {(after prewhitening for 28 frequencies)}. }}
\label{K005988140_Residuals}
\end{center}
\end{figure}

After prewhitening for all the frequencies listed in
Table~\ref{Tab:Freq_Phot}, the final residuals (see
Fig.~\ref{K005988140_Residuals}) do not look
perfectly flat, as one would expect, but reflect small
instrumental effects and reveal discontinuities which were not
accounted for during the homogeneization (pre)processing.

\section{Discussion}

It seems that \citet{Uytterhoeven2011b} based their binary classification on 
the morphology and the high regularity of the light curve (see
Figure~\ref{KIC05988140_LC}), where the deepest minima, occurring
every $\sim$2.91 days, {could be} attributed to primary eclipses. 
Our Fourier analysis {of the Kepler data} reveals the existence of 
multiple frequencies, with the highest amplitude peaks occurring at the 
frequencies of 0.687993~\cd (P$_1$ = 1.45350~d) and 0.343984~\cd (P$_2$ = 2.90711~d), 
{with an exact integer ratio of 2, the first harmonic term being the most significant}. 
A Fourier analysis of the ground-based RV data indicates the same 
frequencies, moreover in the same order of importance, {\it hence showing the double-wave shape}. 
{The fact that a double-wave pattern, almost but not exactly in {anti-}phase with
the light curve (Sect.~\ref{Section:Kplr}), is also detected in the radial
velocity phase diagram excludes binarity as a possible cause of the
observed low-frequency variability. Moreover, the total RV amplitude of 
almost 8 \kms~is too small for a short-period binary system with a primary of 
spectral type A, unless the inclination angle would be very low (in that case, 
we would not detect any feature resembling an eclipse).}  

We {next consider the model} of rotational modulation {generated
by an asymmetric intensity distribution on the stellar surface. The latter
could be (a) due to the presence of stellar spots on the surface, or (b) caused
by a (thin) convection zone located on or near the surface. }  
The RVs folded {on} the period of $\sim$2.91~d (see Figure~\ref{KIC05988140_RV_LC}, 
top) show a double wave which could be explained by a surface structure that 
includes two spots diametrically opposed to one another. The same is also true 
for the double-wave pattern detected in the {\it Kepler} light curve. That a harmonic 
of the rotation period of the star has the highest amplitude in the RV data, was previously 
found in some spotted stars, e.g., in the He-weak silicon star HR\,7224
\citep{Lehmann2006}. On the other hand, both the spectral type (late A)
and the long-term stability of the ``basic'' light curve over a period of about 2~yrs 
argue against variability due to stellar spots for this object.

{Usually, we don't expect magnetic activity (causing spots on the stellar surface)} 
nor convection to occur in the atmospheres of {normal} A-type stars. However, convection 
does occur in some {\it late} A-type stars, and particularly in the photosphere where it 
competes with radiation transporting the flux. The case of Altair is illustrative: this 
fast rotating A-type star has a broad equatorial band at T$_{\rm eff}$ = 6900~K which gives 
rise to strong convection and chromospheric activity like C~II emission~\citep{Peter2006}. 
{Using an XMM-Newton observation, \citet{Robrade2009} confirmed the presence of (coronal) 
X-ray emission and weak magnetic activity.} Altair indeed belongs to the transition region 
where stars can develop envelopes in-between fully radiative and fully convective ones. 
Based on observations collected by the Wide Field Infrared Explorer (WIRE) satellite, \citet{Buzasi2005} 
showed that it is also a very low-amplitude \DSct~star (with $\Delta$m $<$ 1~ppt). 
Interestingly, in addition to the frequencies detected in the 15--29 \cd~range, the existence 
of two frequencies {in the low range, namely} at 3.53 and 2.57 \cd~(cf. their Table~1) is 
demonstrated.
Using long-baseline interferometry, \citet{Peter2006} derived the rotational frequency 
of 2.71 \cd, which is close to the lowest frequency detected by \citet{Buzasi2005}.
In essence, both \citet{Ohishi2004} and \citet{Peter2006} conclude that 
{\it the surface of Altair displays an extremely asymmetric intensity 
distribution and that the asymmetry is consistent with that expected from 
the known high rotation and oblateness.} It thus appears that the rotationally 
distorted stellar surface of Altair 
induces low amplitude modulations which are detectable in the high-quality 
light curves from space. {We can ask ourselves whether KIC\,5988140 might be 
similar to Altair or to the weakly active, X-ray emitting A5/F0-type star HR~8799 \citep{Robrade2010}}. 

\begin{figure}
\begin{center}
\includegraphics[angle=270,scale=0.25]{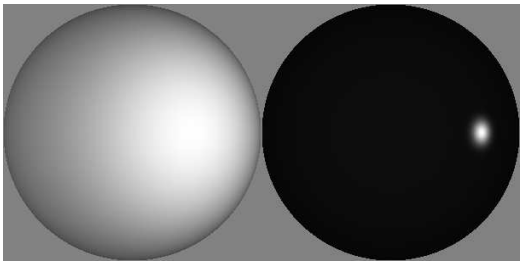}
\caption{{\small Spot models~A (left) and B (right) shown for a certain rotation phase and
drawn on separate greyscales to cover the full brightness range of both cases. }}
\label{KIC5988140_two_models}
\end{center}
\end{figure}

\begin{figure}
\includegraphics[angle=270,scale=0.30]{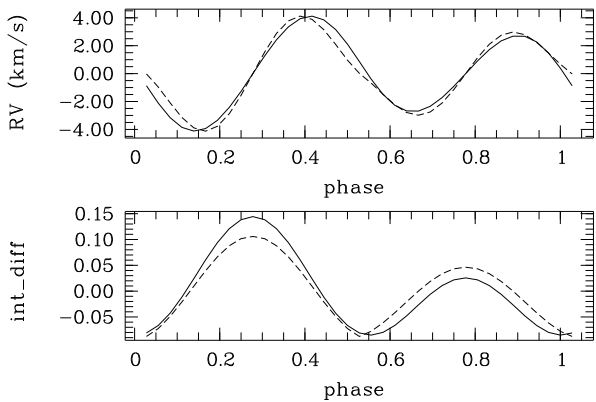}
\caption{{\small RV and relative intensity variations for spot models~A (solid) and B (dashed). }}
\label{KIC5988140_predictions}
\end{figure}

Adopting the radius {of 3.57 $R_{\odot}$} ($R_*$) listed in 
the KIC, we estimate a rotation period of $\sim$3.7~d (reciprocal of 0.27 \cd) 
and of $\sim$2.5~d (reciprocal of 0.40 \cd) for an inclination angle of 
90$^{\circ}$ and 45$^{\circ}$, respectively. The detected period of 2.90711~d 
lies in-between these two derived periods. {We point out that there is 
also a triplet of frequencies ($F_{3,4,5}$)} which has a frequency spacing 
of the order of 0.3~\cd~that could be explained as rotational splitting 
of a single, {independent} frequency. Using $V_{equat} = 50.61 \cdot R_* /P_2 = 
62.3$~\kms~and the measured \vsini\, we obtain an inclination 
angle of $\sim$50\degr, which is a {most reasonable} value. 
{We next attempted to find a model including two hot spots symmetrically 
located on opposite sides of the stellar equator in order to explain the patterns 
in both the light and the RV curves. For this, we used a simple
model in which the stellar surface has intensity $I_{0}$, including a standard
law for the limb darkening, the spots have a size $FWHM_{i}$ (corresponding to the 
Full Width at Half Maximum or FWHM) and a Gaussian intensity distribution of amplitude 
$A_{i}$. The following formula was adopted: \\
\[ I/I_{0} = (1 + spot_1 + spot_2) \cdot limb \]
\[ spot_i = A_i \cdot \exp{[-\ln({2})(r_i/FWHM_i)^{2}]} \]
and
\[ limb = 0.4 + 0.6 \cdot \sqrt{1 - z^{2}}, \]
where $r_{i}$ is the distance on the stellar surface from the centre of spot $i$ and
z is the distance from the centre of the visible disk. For the intrinsic absorption
line profile, we used a Gaussian having a width corresponding to $\xi$\ = 3.16 \kms.
The resulting line profiles are obtained by integrating over the visible surface
where each point has a velocity corresponding to \vsini\, = 52~\kms. Deduced RVs
represent the first moments of the line profiles.

We chose spot sizes and amplitudes such as to reproduce the observed total amplitude 
of the RV curve. Figure~\ref{KIC5988140_two_models} illustrates two cases. In the first case 
(model~A), we chose a maximum size of both spots equal to the stellar radius. An amplitude 
$A_1 = 0.78$ was needed to reproduce the observed RV amplitude of 8 \kms. In the second 
case (model B), we chose a size $FWHM = R_{*}/30$ and needed an amplitude $A_1 = 23$
($A_2$ was chosen a bit smaller in each case to reproduce the observed asymmetry). 
Model~B was (only) selected as the opposite limit case, since such a small and bright spot 
would produce bumps moving across the line profiles (which were not detected). Figure~\ref{KIC5988140_predictions} 
shows the resulting RV and light variations. Both models predict intensity changes 
with a total amplitude of about 20\%, i.e. a factor of 40 larger than the observed one 
(max. 5 ppt peak-to-peak).  
Varying the shape of the spots (exponential instead of Gaussian), their position in latitude, 
their separation in longitude, or the inclination of the rotation axis did not yield a better
agreement. Thus, the small value of the observed light-to-velocity amplitude ratio cannot be 
explained by a (simple) spotted surface model. The same argument also invalidates the assumption 
that KIC\,5988140 might have a partially convective atmosphere (along with the fact that it has 
a low probability of being a fast rotator with \vsini\, = 52~\kms assuming that P$_2$ is the 
rotation period). } 

{Spots will show as regions of different chemical composition on the stellar surface. To verify
this, we performed separate analyses of four {\sc Hermes} spectra taken at four different phases of 
the period P$_2$ (e.g. the extrema of the light curve) to look for possible changes in the derived 
abundances and/or stellar parameters. These results are summarized in Table~\ref{Tab:IndividualAbundances}. 
The largest deviations in \Teff, \lgg, and $\xi$ are respectively 80~K, 0.1~dex and 0.3~\kms, which 
is comparable to the uncertainties listed in Table~\ref{Table:Abundances}. Both the metallicity and 
\vsini~are stable. We may therefore conclude that there is no obvious difference in the obtained 
fundamental parameters nor in the individual abundances with respect to phase. }

\begin{table}\tabcolsep 1.3mm
\caption{Results of the analysis of four individual spectra of KIC\,5988140 taken at four different phases 
of the period P$_{2}$. \Teff~is in K, \vsini~and $\xi$ in \kms, \lgg, [M/H] and the individual abundances 
[A/H] are in dex. The uncertainties are listed in terms of last digits in parentheses.}
\begin{tabular}{llllll}\hline\hline
Param\rule{0pt}{9pt} & Spectr. 1 & Spectr. 2 & Spectr. 3 & Spectr. 4 & Sun\\
\hline \multicolumn{6}{c}{Fundamental parameters}\rule{0pt}{11pt}\\
T$_{{\rm eff}}$\rule{0pt}{11pt} & 7680(50) & 7590(50) & 7630(50) & 7640(50) &\\
$\log{g}$\rule{0pt}{9pt} & 3.58(15) & 3.47(15) & 3.51(15) & 3.57(15) &\\
$v\,\sin{i}$\rule{0pt}{9pt} & 51.0(1.5) & 51.5(1.5) & 50.5(2.0) & 52.0(2.0) &\\
$\xi$\rule{0pt}{9pt} & 3.12(20) & 2.95(25) & 3.30(20) & 3.35(25) &\\
${\rm [M/H]}$\rule{0pt}{9pt} & --0.28(05) & --0.31(05) & --0.30(05) & --0.30(05) &\\
\multicolumn{6}{c}{Individual abundances}\rule{0pt}{11pt}\\
${\rm [Fe/H]}$\rule{0pt}{11pt} & --4.89(05) & --4.94(05) & --4.99(05) & --4.99(05) & --4.59\\
${\rm [Ca/H]}$\rule{0pt}{9pt} & --6.03(10) & --6.03(10) & --6.08(10) & --6.13(10) & --5.73\\
${\rm [Ti/H]}$\rule{0pt}{9pt} & --7.44(10) & --7.44(10) & --7.44(10) & --7.44(10) & --7.14\\
${\rm [Cr/H]}$\rule{0pt}{9pt} & --6.60(15) & --6.65(15) & --6.65(15) & --6.65(15) & --6.40\\
${\rm [Mg/H]}$\rule{0pt}{9pt} & --4.61(15) & --4.61(15) & --4.61(15) & --4.66(15) & --4.51\\
${\rm [Ni/H]}$\rule{0pt}{9pt} & --5.96(15) & --6.01(15) & --6.01(15) & --6.01(15) & --5.81\\
${\rm [Sc/H]}$\rule{0pt}{9pt} & --9.09(20) & --9.19(20) & --9.14(20) & --9.09(20) & --8.99\\
${\rm [Mn/H]}$\rule{0pt}{9pt} & --7.05(20) & --7.10(20) & --7.10(20) & --7.10(20) & --6.65\\
${\rm [V/H]}$\rule{0pt}{9pt} & --8.34(25) & --8.39(25) & --8.39(25) & --8.34(25) & --8.04\\
${\rm [C/H]}$\rule{0pt}{9pt} & --3.75(25) & --3.85(25) & --3.80(25) & --3.75(25) & --3.65\\
${\rm [Y/H]}$\rule{0pt}{9pt} & --9.93(30) & --10.03(30) & --10.03(30) & --9.93(30) & --9.83\\
${\rm [Si/H]}$\rule{0pt}{9pt} & --4.68(30) & --4.78(30) & --4.68(30) & --4.73(30) & --4.53\\
${\rm [Co/H]}$\rule{0pt}{9pt} & --7.47(35) & --7.42(35) & --7.37(35) & --7.42(35) & --7.12\\
${\rm [Sr/H]}$\rule{0pt}{9pt} & --8.92(40) & --8.97(40) & --9.12(40) & --9.17(40) & --9.12\\
${\rm [Ba/H]}$\rule{0pt}{9pt} & --9.27(50) & --9.47(50) & --9.47(50) & --9.37(50) & --9.87\\
${\rm [Eu/H]}$\rule{0pt}{9pt} & -11.52(50) & --11.42(50) & --11.47(50) & --11.57(50) & --11.52\\
\hline
\end{tabular}
\label{Tab:IndividualAbundances}
\end{table}

Third, nine of the photometrically detected frequencies ($F_{3,4,5}$,
$F_7$, $F_{9,10}$, $F_{13}$, $F_{19}$, $F_{23}$, see
Table~\ref{Tab:Freq_Phot}) lie in the typical \DSct\ range. 
{Observed regular spacings are of the order of 0.3 and 0.04~\cd.} The 
question thus remains if the frequencies $F_1$ and $F_2$ {(and the harmonics)} 
can possibly be related to the \GD-like oscillations, {which would confirm its 
classification as a \DSct - \GD\ hybrid star by \citet{Catanzaro2011}.
However, and most importantly, the same reasoning as presented in the model 
involving rotational modulation is applicable here: if the RV variations were caused 
by pulsations, we would expect to detect light variations with a total amplitude 
in scale with the one of the RVs (assuming that low-degree modes only are excited). 
The light-to-velocity amplitude ratio for slowly pulsating B stars lies around 
4-5 mmag/\kms 
\citep[based on 13 confirmed cases, cf. top panel of Fig. 19 in ][]{DC2002}. Assuming 
that this ratio is also valid for \GD\ stars 
\citep[e.g. from Table~2 in ][, we have a mean amplitude ratio of 15 mmag/\kms~from 
two \GD\ stars]{Aerts2004}, 
we would expect a total amplitude of about 16-20 mmag in the ``basic'' {\it Kepler} 
light curve, i.e. 15-20\% (similar to the previous scenario). We are (again) several 
orders below this expected value with the observed total amplitude of the ``basic'' 
light curve (max. 5 ppt). Indeed, the problem of the unbalanced light-to-velocity 
amplitude ratio remains with this explanation too.}  Other, less critical, 
counter-indications are: (1) \GD\ stars are typically multi-periodic pulsators; 
(2) the {small integer} ratio of 2 between the two most dominant frequencies is 
unexpected in the case of (non-radial) pulsators which are not a member of a 
binary system, (3) the existence of more harmonic frequencies, (4) this 
star is perhaps too evolved to be a \GD-type pulsator (cf. its position in 
the H-R diagram), {and (5) the previously derived phase relationship (Sect.~\ref{Section:Kplr}) 
is unusual in the case of pulsation, as the moments of minimum RV (i.e. maximum surface 
layer expansion) generally {\it follow} the moments of light maximum \citep[e.g. in 
$\delta$ Scuti stars, see ][]{Preston1999}.}

\section{Conclusion}

KIC\,5988140 (HD\,188774) was assigned a spectral type of A7.5\,IV-III which,
along with its atmospheric parameters and abundances (see
Tables~\ref{Table:AtmosphericParameters} and
\ref{Table:Abundances}), {puts} it in the middle of the \DSct\ and
{just at} the blue edge of the \GD\ instability strips \citep{Fekel2003}. 
The presence of the multiple frequencies in the range 10--15~\cd~detected
from the {\it Kepler} photometry is firm evidence that the star is
a \DSct\ pulsating star. The {apparently slightly more dominant} low
frequencies detected in both the RV data and in
the {\it Kepler} photometry,  {identified as one major frequency 
with harmonics,} {may be caused by (at least) three very different physical 
processes: a) binarity, b) rotation in combination with an inhomogeneous 
surface intensity/chemical distribution 
or c) \GD-type pulsation in the gravity-mode regime.}  

{In our case, we are left with an open question regarding the phenomenon 
which causes the detected low frequencies in KIC\,5988140 as {\it none} of 
the abovementioned scenarios gives an entirely satisfactory explanation
for the observed variability patterns discovered in this late A-type star. 
It seems indeed difficult to reconcile the $Kepler$ light curve with the ground-based
RV curve. We discarded binarity on the basis of the double-wave
pattern detected in the RV curve. This is a strong argument. 
The light-to-velocity amplitude ratio is an obstacle for the other two scenarios: 
that of rotational modulation as well as that of pulsations. It might be that
our simple modelling of stellar spots is lacking some reality, but we believe 
that it, at least qualitatively, shows the order of the expected effects. }

This reminds us of another still unexplained behaviour observed by {\it Kepler}: 
the fact that the {\it Kepler} light curves of many A- and F-type stars 
show variability at low frequencies \citep{Balona2011}. {The author analysed the 
distributions of \vsini\, derived on the basis of the (dominant) low frequencies
and those of the observed rotational frequencies and showed that both agree rather 
well assuming that the value of the dominant low frequency is the rotational period.} 
{In the specific case of the \DSct\ stars observed by {\it Kepler}, assuming that 
{\it half} of the value of the dominant low frequency is the rotational period, 
\citet{Balona2011} found a good agreement between both (observed and derived) \vsini\, 
distributions. On the other hand, \citet{Grigahcene2010} proposed that most {\it Kepler} 
\DSct\ stars are of the hybrid type where the low frequencies are due to the \GD\ phenomenon. 
An explanation for this behaviour is currently needed. }

KIC\,5988140 is most likely not unique amongst the {\it Kepler} A-type stars. However, it 
is one of the first cases where a high-quality data set in RV has been collected and 
confronted to the photometry. 
{Because of the difficulty in finding a plausible explanation for the cause of the
observed variations, we might consider the more complex model of a triple system to 
explain the double wave detected in the RV curve. We can exclude a hierarchical 
system consisting of a single star and a binary based on the stability criterion because of 
the 1:2 ratio of the long-term frequencies. On the other hand, a single star revolved by two 
satellites could be stable. From the periods and the RV amplitudes, using a typical mass of 
2-3 M$_{\odot}$ for the A-type star, we estimate masses in the range of brown dwarfs to M-dwarfs 
for both objects, namely in the range 20-220 Jupiter masses depending on the orbital inclination angle. 
However, these satellites would have very small semi-major axes of about 8 and 12 R$_\odot$ and we 
cannot tell anything yet about the stability of such close orbits in a 1:2 resonance. From stability 
considerations, the low-mass objects (i.e. brown dwarfs) seem to be more likely. Whether such a 
scenario might also explain the observed photometric variations by reflection (including ellipsoidal 
variability as well as beaming effects), is presently uncertain and will require an in-depth analysis. 
We present this idea merely as a possible working hypothesis for our future work on this 
intriguing $Kepler$ star.}

{One should pay attention to other cases in the $Kepler$ field which might show 
the same difficulty in interpretation. This reminds us that, without complementary
(even ground-based) information in the form of spectroscopy and/or multi-colour 
photometry, any interpretation based on single-passband light curves alone may be 
biased. A valuable method allowing to distinguish between binarity 
rotational modulation or pulsations of type \GD\ would be to collect amplitude 
ratio's in multiple passbands \citep[e.g.][]{Henry2007}. To perform such a task at 
the required scale and accuracy level will however most certainly necessitate 
future space resources. } 

As illustrated by this study, the acquisition of high-quality ground-based spectra 
can provide relevant, new information for a more thorough understanding of 
the hyper-quality white-light stellar photometry provided by the space missions. 

\begin{acknowledgements}
The research leading to these results received funding from the European Research 
Council under the European Community's Seventh Framework Programme (FP7/2007--2013)/ERC 
grant agreement n$^\circ$227224 (PROSPERITY). {Part of this work was also 
supported by the Hungarian grants OTKA K76816, K83790, an E\"otv\"os fellowship, 
and by the J\'anos Bolyai Research Fellowship and the ``Lend\"ulet-2009'' 
Young Researchers Program of the Hungarian Academy of Sciences.}  
{Funding for the {\it Kepler} mission is provided by NASA's Science 
Mission Directorate. We thank the whole team for the development and 
operations of this remarkable mission.} {We furthermore thank Drs. E.~van Aarle 
and B.~Vandenbussche (K.U.Leuven) for help with the acquisition of some of the 
{\sc Hermes} spectra, Drs. V. Antoci and G. Handler for helpful discussions, and 
the referee for valuable comments.} 
This research made use of the SIMBAD database, operated at CDS, Strasbourg, 
France, {and the SAO/NASA Astrophysics Data System}.
\end{acknowledgements}

{}

\appendix
\section{Table A.1 to be published electronically}

\begin{table}[ht] \tabcolsep 0.6mm\caption{\small Journal of recent spectroscopic observations. Listed are the sequence number, the exposure time, the barycentric Julian Date, the measured RV, and the accuracy in RV.}
\label{Table:observations}
\begin{tabular}{rccccc}
\hline\hline
Nr \rule{0pt}{9pt} & \multicolumn{1}{c}{Exp. } & \multicolumn{1}{c}{Date } & \multicolumn{1}{c}{BJD } & \multicolumn{1}{c}{RV } & \multicolumn{1}{c}{Error$_{\rm RV}$}\\
&\multicolumn{1}{c}{(sec)}&&\multicolumn{1}{c}{(2\,455\,000+)}&\multicolumn{1}{c}{(\kms)}&\multicolumn{1}{c}{(\kms)}\\
\hline \multicolumn{6}{c}{{{\sc Hermes}}\rule{0pt}{11pt}}\\
1\rule{0pt}{9pt}& 1200& 2010/06/05 & 353.7043428 &--18.942 & 0.020\\
2&   1800& 2010/07/14 & 392.5180032 &--19.627 & 0.032\\
3&   1800& 2010/07/14 & 392.6444271 &--19.964 & 0.021\\
4&   2000& 2010/07/16 & 394.5169717 &--18.502 & 0.020\\
5&   1800& 2010/07/16 & 394.6163039 &--17.147 & 0.025\\
6&   2700& 2010/07/18 & 396.4641905 &--13.775 & 0.021\\
7&   2400& 2010/07/18 & 396.6385025 &--16.191 & 0.029\\
8&   2700& 2010/07/20 & 398.4383502 &--20.191 & 0.021\\
9&   2400& 2010/07/20 & 398.6246792 &--19.895 & 0.029\\
10&  1200& 2010/08/03 & 412.4096963 &--15.793 & 0.019\\
11&  300 & 2010/08/05 & 414.4434733 &--19.671 & 0.013\\
12&  300 & 2010/08/05 & 414.6630946 &--20.497 & 0.035\\
13&  300 & 2010/08/07 & 416.4173718 &--14.901 & 0.018\\
14&  300 & 2010/08/07 & 416.7144405 &--13.203 & 0.031\\
15&  300 & 2010/08/09 & 418.3981617 &--18.337 & 0.034\\
16&  300 & 2010/08/09 & 418.7080667 &--20.417 & 0.026\\
17&  300 & 2010/08/11 & 420.4033010 &--20.640 & 0.032\\
18&  300 & 2010/08/12 & 421.4271506 &--18.821 & 0.033\\
19&  600 & 2010/08/12 & 421.5680935 &--19.948 & 0.029\\
20&  300 & 2010/08/14 & 423.4211612 &--20.538 & 0.025\\
21&  480 & 2010/08/14 & 423.5536145 &--19.015 & 0.014\\
22&  600 & 2010/08/15 & 424.4009403 &--19.750 & 0.020\\
23&  600 & 2010/08/16 & 425.3884691 &--13.305 & 0.030\\
24&  600 & 2010/08/16 & 425.5342909 &--14.257 & 0.023\\
25&  550 & 2010/08/17 & 426.3884962 &--20.266 & 0.028\\
26&  550 & 2010/08/17 & 426.6004571 &--17.948 & 0.015\\
27&  600 & 2010/08/18 & 427.4912895 &--20.324 & 0.028\\
28&  400 & 2010/08/18 & 427.5044292 &--20.821 & 0.029\\
29&  1800& 2010/08/20 & 429.5096819 &--17.128 & 0.031\\
\multicolumn{6}{c}{{ TLS}\rule{0pt}{11pt}}\\
30\rule{0pt}{9pt} & 900 & 2011/07/12 & 754.5635120 & --19.917 & 0.022\\
31 & 1800 & 2011/07/14 & 757.4733430 & --20.035 & 0.027\\
32 & 1800 & 2011/08/08 & 782.3464050 & --18.128 & 0.033\\
33 & 1800 & 2011/09/10 & 815.4850340 & --18.312 & 0.028\\
34 & 1800 & 2011/09/11 & 816.4960600 & --15.304 & 0.030\\
35 & 1024 & 2011/09/12 & 817.2843500 & --18.059 & 0.032\\
36 & 1800 & 2011/09/12 & 817.4911610 & --14.930 & 0.045\\
37 & 1800 & 2011/09/13 & 818.3643670 & --18.410 & 0.028\\
38 & 1800 & 2011/09/14 & 819.2709620 & --14.222 & 0.020\\
39 & 1800 & 2011/09/14 & 819.2925580 & --14.563 & 0.025\\
40 & 1800 & 2011/09/19 & 824.4577250 & --19.448 & 0.033\\
\hline
\end{tabular}
\end{table}

\end{document}